\documentclass[12pt]{iopart}
\usepackage{amssymb}
\usepackage{graphicx}

\begin{document}

\title{Highly Correlated Electron State and High-Temperature Superconductivity in Iron Pnictides}

\author{M.V.Krasinkova}

\address{Ioffe Physico-Technical Institute, Russian Academy of Sciences, St.Petersburg 194021 Russia}
\ead{marinak@mail.ioffe.ru}

\begin{abstract}
It is shown that the qualitative model of the high-temperature superconductivity suggested earlier for cuprates and doped picene and based on the idea that the valence electron state depends on the character of the chemical bonds they form and on the Coulomb interaction between the electrons is not only confirmed by the experimental data on iron pnictides but is also improved. From the chemical point of view, the high-temperature superconductivity is associated with additional $\pi$ bonding along chains of covalently bonded ions via a delocalized $\pi$ orbital,  just like in cuprates. From the physical point of view, as the data on iron pnictides show, the superconductivity is associated with a FeAs layer transition into the state similar to a macroscopic quantum system characterized by a highly correlated electron state, formation of two-dimensional crystals of electron pairs with quantized energy levels, and a strong Coulomb interaction between these crystals. Superconductivity in such a system is accomplished by a two-dimensional Wigner crystal consisting of one-dimensional Wigner crystals formed by bosons, i.e., singlet electron pairs that are in the same quasi-one-dimensional state extending along the ion chain, which corresponds to a delocalized $\pi$ orbital in chemistry. The model applicability to three different classes of materials (cuprates, picene, iron pnictides) indicates that it can prove useful for development of the theory of superconductivity taking into consideraion the highly correlated state of valence electrons and strong Coulomb interactions between the electrons.

\end{abstract}

      \section{Introduction}
\label{se:intro}
      
Discovery of a new family of high-temperature superconductors based on iron pnictides and chalcogenides \cite{ref1,ref2,ref3,ref4} attracted considerable attention of researchers. It was expected that comparison of properties of two families of superconducting materials (based on copper and iron) could get an insight into the mechanism of carrier pairing and reveal basic interactions responsible for superconductivity at rather high temperatures. 

The research efforts have already given results which lead to some interesting conclusions. For instance, the highly contradictory data on the isotopic effect magnitude and sign in iron-based superconductors \cite{ref5,ref6,ref7} show that electron-phonon interactions are unlikely to be responsible for the carrier pairing mechanism. This is confirmed by a zero isotopic effect in cuprates demonstrating the highest Tc \cite{ref8}. 

The absence of magnetic ordering in some iron-based superconducting materials \cite{ref9,ref10} and also a very low magnetic moment per Fe$^{2+}$ in the range from 0.25 to 0.36 $\mu_B$ \cite{ref11,ref12} which is almost an order of magnitude lower than the moment of 2.3 $\mu_B$ predicted theoretically \cite{ref13} for iron-based superconductors cast doubt on another explanation of the pairing mechanism suggested earlier for cuprates and including participation of short-range spin fluctuations \cite{ref14}. 

However, the most intriguing finding of the comparative analysis is a number of specific features shared by iron- and copper-based superconductors. Both families are characterized by 
\begin{enumerate}
\item a layered structure consisting of alternating conducting and insulating layers with ionic bonding between them, 
\item a charge transfer from insulating layers into a conducting layer and, hence, a charged state of the conducting layer, 
\item a correlation between superconductivity and crystal lattice distortions, 
\item a bell-shaped dependence of Tc on dopant concentration, 
\item chemical pressure of the layers surrounding the superconducting layer and the dependence of Tc on this pressure, 
\item a high second critical field Hc$_2$, 
\item the presence of anions with filled outer electron inert-gas shells and cations with unoccupied electron states.
\end{enumerate}

The similarities between the two families raise a question as to whether the qualitative superconductivity model suggested earlier for cuprate superconductors \cite{ref15,ref16} is applicable to the new materials. The paper is devoted to finding an answer to this question.

\section{Specific features of chemical bonding in multicomponent compounds based on iron pnictides}
\label{se:2}

The idea underlying the high-Tc superconductivity mo\-del for cuprates \cite{ref15,ref16} is that characteristic properties of a material originate from specific features of chemical bonding because the character of bonding determines the state of valence electrons participating in it. Therefore, to define the electron state of the compounds based on iron pnictides, the chemical bond character should be first of all considered. This makes the problem interdisciplinary (physics and chemistry).

A small difference between electronegativities of iron and arsenic (less than $\sim 0.25$) indicates that the bonding between these elements must be purely covalent. It is confirmed by the binary compound FeAs$_2$ characterized by a marcasite structure which is typical of covalent compounds. However, a charge transfer into the FeAs layer from neighboring layers in the multicomponent compounds can change the character of bonding between elements in the layer. This is evidenced by a change in the iron coordination number from 6 in the binary compound to 4 in the FeAs layer, and it can be supposed that bonding between Fe and As in the multicomponent compounds is a mixed ionic-covalent bonding rather than covalent one. Indeed, introduction of additional electrons into the FeAs layer converts an arsenic atom into a negatively charged As$^{1-}$ ion, the outer electron shell of which is similar to that of an oxygen atom ($s^2p^4$). Owing to this, As$^{1-}$, like an oxygen atom, oxidizes iron to the Fe$^{2+}$ state and is transformed into As$^{3-}$ characterized by a fully occupied outer electron shell ($4s^24p^6$). It would seem that As$^{3-}$ must prefer purely ionic bonding but this is prevented by polarization of an anion by a cation which makes the bonding mixed (to be discussed in detail below). It is also important to note that As$^{3-}$ outer electrons are weakly bound with this ion, which is evidenced by a very large ionic radius of As$^{3-}$ (2.22 \AA), well above the atomic radius (1.39 \AA).

Thus, a negatively charged (owing to charge transfer) FeAs layer in the multicomponent compound based on iron pnictides consists of Fe$^{2+}$ and As$^{3-}$ ions with a mixed ionic-covalent bonding between them rather than of Fe and As atoms covalently bonded with each other. This fact is extremely important for understanding properties of the multicomponent compounds because of several factors:
\begin{enumerate}
\item ionic compounds are characterized by ordering of local distortions which lowers the deformation energy of the ionic quasi-two-dimensional (quasi-2D) lattice of the FeAs layer and promotes formation of quasi-one-dimensional (quasi-1D) or one-dimensional (1D) structures (or ion chains);  
\item a strong electric field generated by As$^{3-}$ around Fe$^{2+}$ can cause a Fe$^{2+}$ low-spin state accompanied by localization of d-electrons at its inner electron shell and thus block their participation in chemical bonding between ions. (The low-spin state of Fe$^{2+}$ was discussed for the FeP layers characterized by the crystal structure identical to that of the FeAs layers \cite{ref17} and for LaFeAsO$_{0.89}$F$_{0.11}$ \cite{ref10});
\item a similarity between FeAs layers in pnictides and CuO$_2$ layers in cuprates in which charge transfer also occurs and which also contain cations with small ionic radii and anions with rather large ionic radii becomes more obvious.
\end{enumerate}

The similarity suggests that cations in the FeAs layer also polarize anions, attract their weakly bound electrons and share them in the case of overlapping of ion electron states. The electron sharing by ions means that covalent bonds or, to be more exact, covalent components of mixed ionic-covalent bonds are formed between the ions, and the bonds are localized due to the polarization of As$^{3-}$ by Fe$^{2+}$.

Let us consider now what covalent bonds can be formed between Fe$^{2+}$ and As$^{3-}$ in the FeAs layer consisting of FeAs$_4$ tetrahedra.  

Fe$^{2+}$ with its outer electron $3d^64s^0$ shell can form four tetrahedrally oriented bonds (Fig.1a) only if it passes into the $sp^3$ hybrid state which is formed by using unoccupied $4p$ states. Such hybridization keeps up a low-spin state with a symmetric $d^6$ shell, when $d$-electrons do not take part in chemical bonding and the Fe$^{2+}$ participation in covalent bonding is limited to making unoccupied hybrid $sp^3$ states available.

\begin{figure*}
  \includegraphics{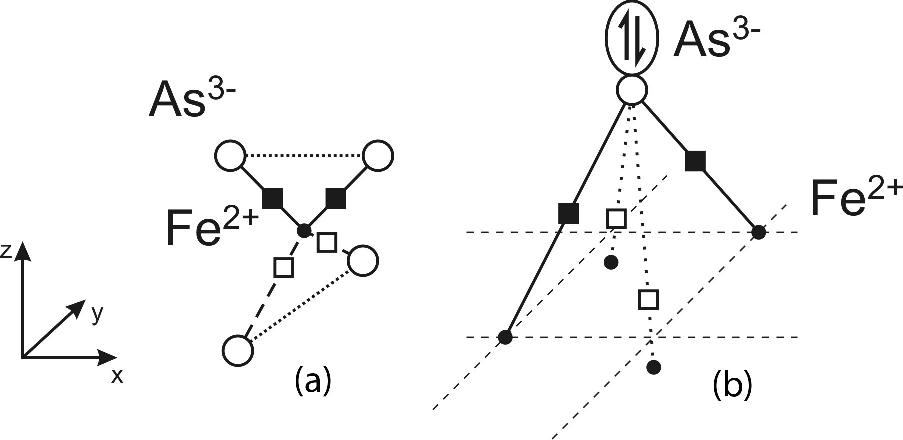}
\caption{Coordination polyhedra of Fe$^{2+}$(a) and As$^{3-}$(b). Different lengths of 1e (dashed lines in Fig.1a and dotted lines in Fig.1b) and 2e (solid lines) bonds and one of two types of distortions of the As$^{3-}$ pyramid base plane. Black squares show electron pairs, empty squares show unpaired electrons. The double arrow shows an unshared electron pair in the apical As$^{3-}$ state}
\label{fig:1}       
\end{figure*}

Concerning As$^{3-}$ and its ability to form four bonds with four Fe$^{2+}$ lying approximately in one plane (Fig.1b), its hybrid state must be either $d^4s$ or $d^4p$, where $d$ is the unoccupied $4d$ states of As$^{3-}$. This hybridization gives rise to the As$^{3-}$ coordination polyhedron in the form of a square pyramid typical of this ion in the compounds considered. According to Hund's rule, eight valence electrons of As$^{3-}$ (its own two $s$- and three $p$-electrons, two electrons from the iron atom and one electron transferred from another layer) distributed among five hybrid states must form three singlet electron pairs that occupy three states and one triplet pair that occupies two remaining states. In the case of such an occupation of the states, As$^{3-}$ can form two usual two-electron ($2e$) bonds (by using two singlet electron pairs) and two one-electron (1e) bonds (by using a triplet pair) with four Fe$^{2+}$ ions. Thus, covalent bonds in the As$^{3-}$ coordination polyhedron prove to be nonequivalent in energy and, hence, in length. (The presence of bonds of different lengths was experimentally confirmed by observation of two Fe-As distances \cite{ref12,ref18}). A third singlet electron pair of As$^{3-}$ must, in all probability, occupy the state at the square pyramid apex and remain an unshared pair, i.e., not participating in covalent bonding (Fig.1b). It takes part in ionic bonding between the FeAs layer and neighboring layers with another chemical composition. 

Formation of nonequivalent polarization-localized bonds disturbs spherically symmetric electron density distribution in the As$^{3-}$ shell (typical of a fully occupied shell). Thus, the mixed ionic-covalent chemical bonding in the FeAs layers results in a highly nonuniform electron density distribution in the coordination polyhedra of ions.

\section{Ordering of nonequivalent chemical bonds, formation of ion chains with identical bonds, and splitting of the Fe$^{2+}$ plane into two planes}
\label{se:3}
Formation of nonequivalent bonds between Fe$^{2+}$ and As$^{3-}$ must be accompanied by distortions in the coordination polyhedra of the ions. This is seen, first of all, in the coordination polyhedron AsFe$_4$ of As$^{3-}$, in which the nonequivalent bonds form lateral edges. Because of a dense packing of large As$^{3-}$ ions that prevents their displacement, their polyhedra with edges of different lengths can be formed only due to Fe$^{2+}$ displacements from their positions in the plane of an ideal square pyramid base (Fig.1b). These displacements are possible due to rather large tetrahedral cavities formed by densely packed As$^{3-}$ ions. However, because of a strong Coulomb interaction between the localized unpaired electrons (i.e., the electrons that form 1e bonds) and between electron pairs in neighboring Fe$^{2+}$ coordination polyhedra, these Fe$^{2+}$ displacements cannot be arbitrary and uncorrelated in neighboring tetrahedra in the FeAs layer. The most favorable type of displacement that reduces the Coulomb interaction between electrons of neighboring tetrahedra is the Fe$^{2+}$ displacement from the tetrahedron center towards its two unshared edges (Fig.2). This displacement contributes to formation of two short and two long bonds in each FeAs$_4$ tetrahedron and AsFe$_4$ square pyramid (Fig.1) and is accompanied by deviations of the angles between bonds from the ideal tetrahedron angle ($\beta_{id}=109.47^\circ$), i.e., an increase in the angle between 2e bonds ($\beta_1 > \beta_{id}$) and a decrease in the angle between 1e bonds ($\beta_2 < \beta_{id}$) (Fig.2). The observation of such deviations \cite{ref12} can be regarded as one of the proofs of existence of Fe$^{2+}$ displacements in their tetrahedra, though this experimental fact has not been associated so far with the idea of Fe$^{2+}$ displacement. 

\begin{figure*}
  \includegraphics[width=0.75\textwidth]{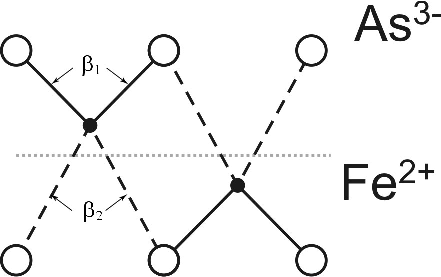}
\caption{Schematic representation of Fe$^{2+}$ displacement from the centre of an ideal tetrahedron (the dotted line shows the level of the centre) in two neighboring FeAs$_4$ tetrahedra and formation of 1e and 2e bonds (dashed and solid lines, respectively)}
\label{fig:2}       
\end{figure*}

The Fe$^{2+}$ displacements in neighboring tetrahedra in positive and negative directions of the Z axis (Fig.2) result in formation of two Fe$^{2+}$ planes instead of one plane typically considered. In other words, the Fe$^{2+}$ plane in the FeAs layer is divided into two planes separated by the distance depending on the difference in the nonequivalent bond lengths and their angles of inclination to the Z axis. Note that the FeAs layer itself is divided by the two Fe$^{2+}$ planes into two parts (or two halves).

Two variants of Fe$^{2+}$ ordered displacement in neighboring tetrahedra in the FeAs layer are possible, i.e., alternating along both the X and Y axes (in the chess-board order in the XY plane) or only along one of these axes (Fig.3). In the first case the interionic Fe$^{2+}$-Fe$^{2+}$ distance in the As$^{3-}$ coordination pyramid base remains the same between all Fe$^{2+}$, while in the second case there must be two different distances. Therefore, the observation of two different Fe-Fe distances for undoped LaOFeAs \cite{ref12} and for the orthorhombic phase of CaFe$_2$As$_2$ \cite{ref19} can be regarded as a direct confirmation of ordered Fe$^{2+}$ displacement in these materials. The differences of $\sim$ 0.015 {\AA} and $\sim$0.037 {\AA} in the Fe$^{2+}$-Fe$^{2+}$ distances obtained from the experimental data of \cite{ref12} and \cite{ref19}, respectively, can be used to roughly estimate the ion displacement along the Z axis and the interplane distance. The estimation yields a displacement of $\sim$0.14 {\AA} and $\sim$0.22 {\AA} and a distance of $\sim$ 0.28 {\AA} and $\sim$ 0.44 {\AA} for LaOFeAs and CaFe$_2$As$_2$, respectively.

\begin{figure*}
  \includegraphics[width=0.75\textwidth]{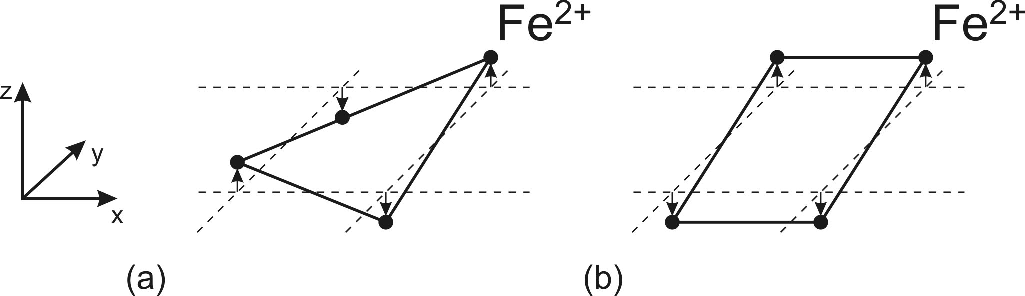}
\caption{Two types of distortion of the AsFe$_4$ square pyramid base resulting from Fe$^{2+}$ ordered displacements in the tetragonal (a) and orthorhombic (b) phases}
\label{fig:3}       
\end{figure*}

The Fe$^{2+}$ displacement ordering is accompanied by ordering of nonequivalent bonds in the FeAs layer. In the first case (i.e., displacement in the chess-board order), identical bonds are ordered in the FeAs layer as quasi-1D ion chains oriented in the direction of one of two diagonals in the square pyramid base (Fig.4b) (The chain quasi-one-dimensionality is due to the chain zig-zag contour along the Z axis determined by the As$^{3-}$ coordination pyramid height above the Fe$^{2+}$ planes). This bond ordering is typical of the tetragonal phase for which there must be no difference in the Fe$^{2+}$-Fe$^{2+}$ distances, consistent with experimental observations \cite{ref19}. In the second case, ion chains bound by identical bonds become zig-zag-like not only along the Z axis, but also in the XY plane (Fig.4a). This bond ordering is typical of an orthorhombic phase. Thus, both types of ordering lead to formation of ion chains with identical bonds, which means that the quasi-2D crystal lattice of the FeAs layer is separated into quasi-1D ion chains.

\begin{figure*}
  \includegraphics[width=0.75\textwidth]{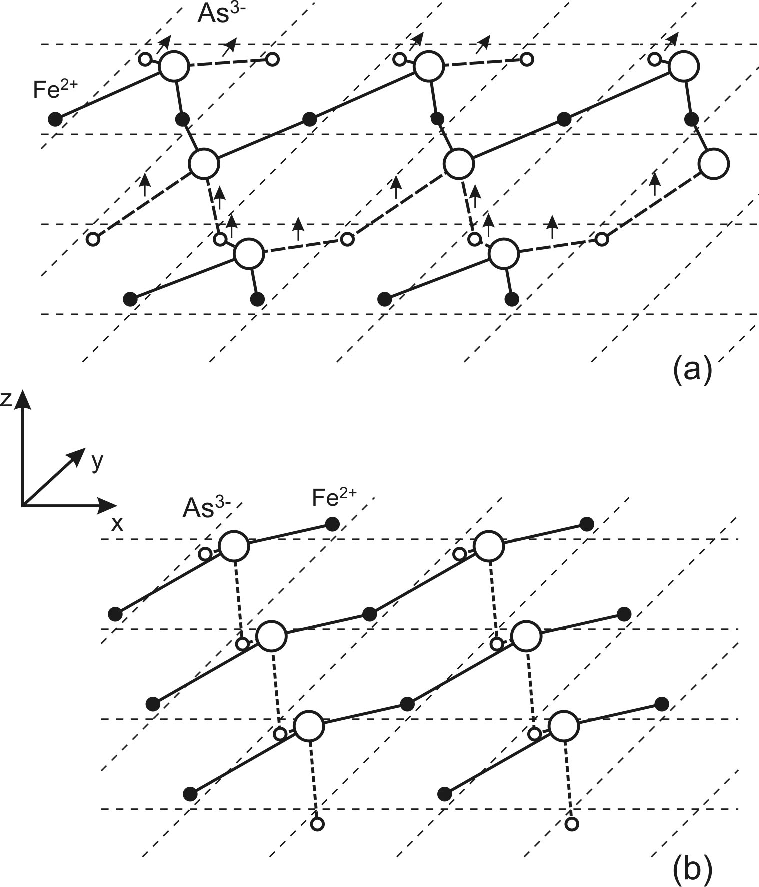}
\caption{Ion chains bound by 2e bonds (solid lines) and separated from each other by static 1e bonds (dashed lines) in the orthorhombic phase (a) and by resonating 1e bonds (dotted lines) in the high-temperature tetragonal phase (b). The chains are shown for one half of the FeAs layer. The arrows show spins of unpaired electrons of 1e bonds which are ferromagnetically ordered in each chain. Small empty circles show Fe$^{2+}$ belonging to the other half of the FeAs layer}
\label{fig:4}       
\end{figure*}

\section{The state of valence electrons bound to ions and interacting with each other and a sandwich-type charge distribution in the FeAs layer}
\label{se:4}
Since the compounds based on iron pnictides are characterized by chemical bonds of different characters (pu\-re\-ly ionic and mixed ionic-covalent, the mixed bonds being nonequivalent, i.e., 1e resonating or nonresonating and 2e bonds), the states of valence electrons forming all the bonds in the pnictide layer cannot be the same. Three main states can be separated out: the state of singlet pairs; the state of unpaired electrons; and the state of unshared singlet pairs localized at anions and participating in ionic bonding between pnictide layers and layers with other compositions. Note that there can also be another state of unshared pairs typical of the pairs formed from unpaired electrons in the low-temperature tetragonal phase (see Sect. 6 and 7).

Let us consider in detail each of the electron states in the FeAs layer to understand the general picture of the behavior of valence electrons characterized by strong interactions between nearly all electrons. 

The electrons of 2e bonds form singlet pairs and are in the hybrid electron state resulting from overlapping of As$^{3-}$ ($d^4s$) and Fe$^{2+}$ ($sp^3$) hybrid states. The singlet pairs are localized by anion-cation polarization in the space between ions and are distributed in a specific pattern that depends on both the 2e bond ordering in the ion coordination polyhedra and Coulomb interaction between these neighboring electron pairs. As a result, spatial arrangements of the electron pairs for tetragonal and orthorhombic phases are different. However, in general, the states of the pairs in both phases can be regarded as the state of a quasi-2D crystal of electron pairs localized in the volume of a quasi-2D ionic crystal which is the FeAs layer. (The prefix "quasi" is due to a definite layer thickness, about 8 \AA). Since the FeAs layer is divided into two parts by two Fe$^{2+}$ planes, it is convenient to think of the state of electron pairs as two 2D crystals of electron pairs each of which is situated in one half of the layer. Moreover, owing to formation of 1D or quasi-1D chains of identical bonds (Sect. 3), each of these two 2D crystals of electron pairs proves to be consisting of 1D or quasi-1D crystals which belong to different chains (Fig.5a, c). Because As$^{3-}$ are arranged along the Z axis in the chess-board order with respect to the Fe$^{2+}$ planes, the 1D crystals of pairs in two FeAs layer parts are displaced at half of the inter-ion distance (in one half of the layer) relative to each other.

\begin{figure*}
  \includegraphics[width=0.75\textwidth]{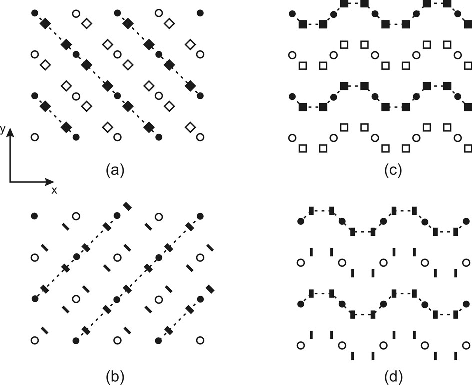}
\caption{Projections (on the XY plane) of two 2D electron crystals of electron pairs forming 2e bonds (a) and two 2D electron crystals of unpaired electrons forming 1e bonds (b) in both parts of the FeAs layer together with the projections of two Fe$^{2+}$ planes for the tetragonal phase. The thick dashes and black squares and circles show unpaired electrons, electron pairs, and Fe$^{2+}$ ions, respectively, which are in one part of the layer. The thin dashes, empty squares and circles show the same for the other part of the layer. Similar schematic pictures are given for the orthorhombic phase (c, d). The dotted lines are aids for the eye. Distances between the 1D and quasi-1D electron crystals in each part of the FeAs layer are 3.9 {\AA} for (a, b) and 5.6 {\AA} for (c, d). As$^{5+}$ are not shown to make the figure simpler}
\label{fig:5}       
\end{figure*}

The main function of the electron crystals of pairs is formation of 2e covalent bonds between ions which are the strongest bonds in the FeAs layer and, hence, determine its crystal structure. For this reason, the state of these pairs (or, to be precise, the electron crystal of these pairs) should be regarded as strongly bound to ions (simultaneously to As$^{3-}$ and Fe$^{2+}$).

The valence electrons that form 1e bonds (because of this they are regarded as unpaired) are also in the hybrid electron state formed by overlapping of hybrid ion states. However, the degree of overlapping in this case is much lower than that for 2e bonds considered above. This means that bonding of the unpaired electrons with ions is weaker. The unpaired electrons are also localized by polarization, strongly interact with each other, and form a periodic quasi-2D electron lattice separated by Fe$^{2+}$ planes into two 2D electron crystals, each of which, in its turn, is divided into 1D or quasi-1D crystals (due to formation of ion chains) (Fig.5b,d). The spatial pattern of unpaired electrons is similar to that of electron pairs for both phases. However, the crystals of unpaired electrons and electron pairs are displaced in the FeAs layer with respect to each other both along the Z axis and in the XY plane because of different bond lengths and angles between the bonds formed by the electrons and electron pairs. The mutual orientation of the crystals in the XY plane for tetragonal and orthorhombic phases is shown in (Fig.5a-b and c-d, respectively).

One more specific feature of the unpaired electrons should be pointed out. We regard them as unpaired only because they form 1e bonds. Actually, every two unpaired electrons form a triplet pair if they occupy two identical electron states in one As$^{3-}$ coordination polyhedron. Therefore, a crystal of unpaired electrons can also be referred to as a crystal of triplet pairs. The first presentation is convenient for understanding the general picture of electron-electron interactions in the FeAs layer, the second one is useful for understanding physical properties of a material (metal-like conductivity and magnetic ordering). 

The basic function of crystals of unpaired electrons, as well as crystals of pairs, is formation of additional covalent bonds between ions. However, in contrast to electron pairs, unpaired electrons are weakly bound to As$^{3-}$ and Fe$^{2+}$. This is evidenced by a low energy of a 1e bond which can be estimated from the magnetic ordering temperature $T_N$ because magnetic ordering in materials with mixed chemical bonding is associated with 1e bond formation \cite{ref15,ref16}. Estimates show that this energy is about two orders of magnitude lower than the 2e bond energy. A weak bonding of unpaired electrons with ions is caused by Coulomb repulsion of the electrons from electron pairs in the ion coordination polyhedron (Fig.1) or, to be more exact, Coulomb repulsion between crystals of unpaired electrons and crystals of electron pairs. The dependence of the state of unpaired electrons on Coulomb interaction inside the ion coordination polyhedron makes it very sensitive to even minor polyhedron deformations. Thus, unpaired electrons (or crystals of unpaired electrons) are the most weakly bound to ions in the FeAs layer and their state is the most susceptible to polyhedron deformations, i.e., they will be the first to change their state under external actions.

The third group of valence electrons includes unshared singlet electron pairs localized in apical hybrid As$^{3-}$ states and, hence, strongly bound to As$^{3-}$. Owing to Coulomb repulsion between pairs at neighboring As$^{3-}$, these pairs form 2D crystals of unshared electron pairs that can be regarded, in all probability, as Wigner crystals of unshared pairs. The Wigner crystals are localized and form negatively charged surface planes in the FeAs layer, and their role is to provide purely ionic bonds between the FeAs layer and neighboring positively charged layers with another chemical composition (e.g., LaO$^{1+}$, Ca$^{2+}$, Ba$^{2+}$).

Note that we refer to the electron crystal of unshared pairs in this manuscript as a "Wigner crystal" (suggested for electrons)  instead of introducing a new notion because the conditions of formation of our crystal  (the distance between pairs is much longer than the Bohr radius, localization of the unshared pairs under the action of Coulomb repulsion) and its behavior (ability to be delocalized as a whole without changing its energy) are similar to those of the Wigner crystal of electrons. Moreover, delocalization of the Wigner crystal of electrons gives rise to the Fr\"{o}hlich conductivity, and  delocalization of the Wigner crystal of pairs, as wil be shown below, gives rise to the superconductivity. Therefore, the extension of the well known notion of a "Wigner crystal" to the electron crystal of unshared pairs seems quite reasonable.

This analysis of the valence electron state in the FeAs layer shows that in spite of the fact that electrons are divided into three groups in accordance with the character of the bonds they form, they have much in common, i.e., (i) all of them are characterized by localization and bonding with ions (As$^{3-}$ or As$^{3-}$ and Fe$^{2+}$ together) which can be strong (electron pairs or unshared pairs) and weak (unpaired electrons), (ii) electrons in each of the groups strongly interact with each other and form electron crystals, and (iii) electron crystals of different groups also interact with each other. All the Coulomb interactions result in ordering of all charges in the FeAs layer (including ion charges) with some periodicities in all crystallographic directions. The projections of electron crystals and Fe$^{2+}$  planes on the XY plane are shown in Fig.5. The charge distribution along the Z axis can be pictured as a sandwich structure consisting of 10 charged planes which include 6 planes of 2D electron crystals and 4 planes of positively charged ions (Fig. 6a). The half of this sandwich structure consists of two 2D electron crystals of electron pairs (unshared anion pairs and pairs of 2e bonds), a 2D crystal of unpaired electrons, a plane formed by Fe$^{2+}$, and a plane formed by As$^{5+}$. (Here the arsenic ion is considered as As$^{5+}$ rather than As$^{3-}$ because we consider its valence electrons as forming their own separate planes, i.e., planes of electron crystals). The specific feature of the sandwich structure is very small inter-plane distances along the Z axis (of the order of tenths of \AA), which is an order of magnitude lower than the minimal inter-charge distances in the planes (it is about 3.9 {\AA} in the Fe$^{2+}$ and As$^{5+}$ planes, 1.65 {\AA} in the crystal of unpaired electrons, and 1.96 {\AA} in the crystal of electron pairs). This suggests that there should be rather a strong Coulomb interaction between all charged planes along the Z axis. Besides, due to large inter-charge distances in the planes and relative charge displacements in the XY plane in both parts of the FeAs layer, electron crystals can experience Coulomb interaction with not only the neighboring electron crystals in their half of the FeAs layer but also with the electron crystals of the second layer half. For instance, the Fe$^{2+}$ planes having the largest ion-to-ion distance do not inhibit Coulomb interaction between unpaired electrons or electron pairs of one part of the FeAs layer with those of the second part of the FeAs layer (to be discussed below).

\begin{figure*}
  \includegraphics[width=0.75\textwidth]{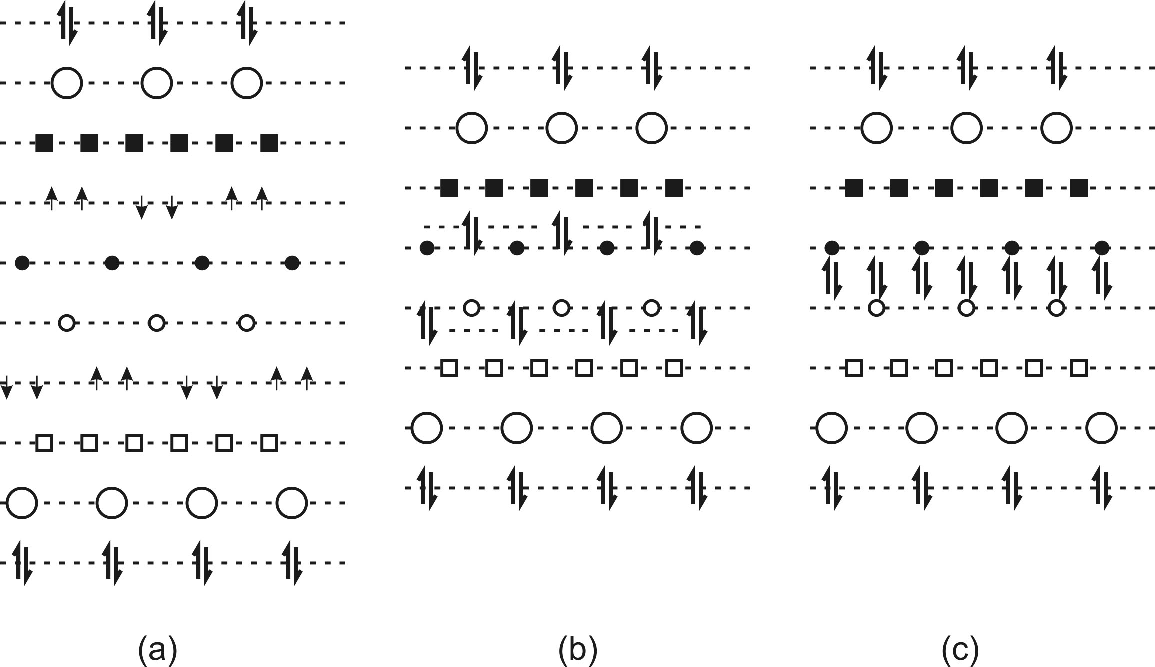}
\caption{A sandwich-type charge distribution along the Z axis in the FeAs layer and its change with increasing Coulomb interaction between electron crystals. The sequence of the planes in (a) is as follows (from top to bottom): 2D crystal of unshared pairs, As$^{5+}$ plane, 2D crystal of pairs of 2e bonds, 2D crystal of unpaired electrons (or triplet pairs), Fe$^{2+}$  plane. The order of planes is reverse in the second part of the layer. The designations are the same as in Fig.5. Large empty circles show As$^{5+}$.  Fig.6a shows the charge distribution in the FeAs layer in the orthorhombic phase (without doping and pressure application) characterized by two 2D Wigner crystals of unpaired electrons. Fig.6b shows an intermediate, probably, unstable state characterized by two 2D Wigner crystals of unshared pairs shown by double arrows near the Fe$^{2+}$ planes. Fig.6c shows the superconducting state characterized by the united 2D boson Wigner crystal situated between two Fe$^{2+}$ planes}
\label{fig:6}       
\end{figure*}

\section{Conservation of electron crystal of unpaired electrons at high temperatures and the origin of its spin ordering}
\label{se:5}

We consider the temperature dependence of the state of valence electrons in undoped materials for the most studied LaOFeAs compound as an example. It has been experimentally shown that LaOFeAs is at room temperature in a high-temperature tetragonal phase. As temperature decreases to T$_N=170K$, a transition to an orthorhombic phase is observed, which points to a change in chemical bonding between ions. The change must be mainly associated with 1e bonds, which are the weakest and easily broken by thermal lattice vibrations. It is believed that above T$_N$ they can exist only in a dynamic state, i.e., in resonance with ionic bonds, and pass to a static state below $T_N$. The state of the unpaired electrons that take part in the resonating bonds must be intermediate between the state of their localization in the space between ions (at formation of 1e bond) and the state of their localization at As$^{3-}$ (at ionic bonding). So, the localized state of unpaired electrons in both cases is conserved and the electron crystal state is also conserved because it results from Coulomb interaction between the electrons at their neighboring localization sites. This means that unpaired electrons in parent (undoped) materials are in the state of electron crystals at rather high temperatures even though they are weakly bound to ions and form only resonating bonds. Note also that the resonance bond state in the FeAs layer is, probably, realized via Fe$^{2+}$ thermal vibrations which (as shown above) are highly anisotropic and have the maximum amplitude in the plane of 1e bonds which is orthogonal to the 2e bond plane (Fig. 1a).

Since the transition into the orthorhombic phase at T$_N$ is associated with formation of quasi-1D chains of ions bonded to each other by static 1e bonds (Fig.4a), it is accompanied by arising exchange interaction between electrons of triplet pairs of neighboring As$^{3-}$ when the electrons of two neighboring As$^{3-}$ form two 1e bonds with one and the same Fe$^{2+}$. The exchange interaction results in ferromagnetic spin ordering of all electrons in such quasi-1D ion chains, i.e., ferromagnetic ordering of quasi-1D crystals of the unpaired electrons. (Just this ordering was observed experimentally \cite{ref10,ref11,ref12}). Thus, the ferromagnetic ordering of quasi-1D electron crystals is determined by both triplet electron pairing within the coordination polyhedron of each As$^{3-}$ and triplet pairing between electrons belonging to two neighboring As$^{3-}$ polyhedra in the chain (at formation of 1e bonds with one and the same Fe$^{2+}$). 

A purely magnetic interaction between spin-ordered quasi-1D electron crystals must make the 2D crystals of unpaired electrons they form antiferromagnetically ordered. Since magnetic interaction is much weaker than the exchange interaction responsible for ferromagnetical ordering, the total magnetic moment per Fe$^{2+}$ is bound to be low. The antiferromagnetic ordering of neighboring ferromagnetically ordered chains and a low magnetic moment per Fe$^{2+}$ in FeAs layers were observed experimentally \cite{ref10,ref11,ref12}.

It follows from the above discussion of magnetic ordering in the FeAs layers that the magnetic moment determination on the one-iron-ion basis is not quite correct, the more so that Fe$^{2+}$ in the FeAs layer, as noted above, is in the low-spin state and, hence, has no its own magnetic moment. The magnetic moment characterizes in this case the entire 2D electron crystal consisting of quasi-1D ferromagnetically ordered electron crystals.

Because magnetic ordering in the FeAs layers is associated with spin ordering of electron crystals, it must collapse when spin ordering collapses (e.g., at transition to the resonance state of 1e bonds above T$_N$ or at pairing of unpaired electrons). This means that magnetism and superconductivity cannot coexist in the compounds based on iron pnictides because superconducting pairs in them, as will be shown below, are formed by pairing of the unpaired electrons.

\section{Singlet pairing of unpaired electrons and formation of Wigner crystals of pairs under doping}
\label{se:6}
Analysis of chemical bonding in the iron-based compounds has shown that doping in these compounds, like in cuprates, is bound to be accompanied by both the formation of unoccupied Fe$^{2+}$ electron states and a change in the state of unpaired electrons. Besides, the state of unpaired electrons is bound to change drastically: first, the change must embrace all the unpaired electrons, and, second, it must involve transition from triplet to singlet electron pairing.

Let us consider in detail changes in the unpaired electron state for LaOFeAs doped with fluorine. A partial substitution ($\sim$ 10 percent) of oxygen by fluorine is accompanied by introduction of additional electrons into the FeAs layer. It can be supposed that an introduced electron is paired with one of two unpaired electrons of As$^{3-}$ that participate in formation of 1e bonds with Fe$^{2+}$. The pair formation breaks the 1e bond, bonding between the ions becomes purely ionic, and Fe$^{2+}$ gets an unoccupied electron state. Transition to ionic bonding (instead of formation of a new 2e bond) results from Coulomb repulsion of the pair from two electron pairs of 2e bonds in the Fe$^{2+}$ coordination polyhedron (Fig.1a). For this reason the singlet pair formed localizes at As$^{3-}$. Its bonding with As$^{3-}$ becomes much weaker due to an increase in Coulomb repulsion of the newly formed pair from the 2e bond pairs already present in the As$^{3-}$ coordination polyhedron (Figs.1b and 7). It is quite possible that the new pair will even pass from the $d^4s$ state into a higher-energy As$^{3-}$ state formed by adding a part of the $p$ state (or even into the $d^4p$ state). This pair can be regarded as an unshared electron pair of As$^{3-}$ (in addition to the unshared pair in its apical electron state). Thus, introduction of electrons at doping leads to formation of unshared electron pairs in the As$^{3-}$ electron shell which prove to be weakly bound to As$^{3-}$. 

\begin{figure*}
  \includegraphics[width=0.75\textwidth]{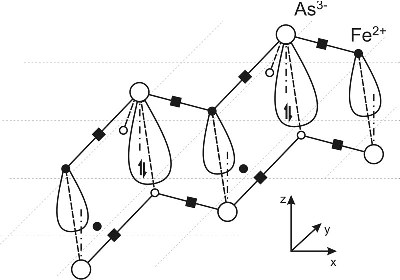}
\caption{Mutual arrangement of electron states of unshared pairs and Fe$^{2+}$ unoccupied states and their alternation along the chain of covalently bonded ions, and also the position of the unshared pair (double arrow) with respect to the electron pairs in the coordination polyhedron of its As$^{3-}$ and two pairs of nearest neighboring As$^{3-}$ situated in the second part of the FeAs layer (All the pairs are shown by black squares)}
\label{fig:7}       
\end{figure*}

However, formation of singlet electron pairs in the As$^{3-}$ electron shells which have got additional electrons does not terminate the process of change in the state of unpaired electrons. Because the initial state of unpaired electrons in undoped materials is characterized by the state of 2D electron crystal, formation of a singlet pair (instead of an unpaired electron) in this electron crystal will be accompanied by the enhanced Coulomb repulsion between this pair and the nearest unpaired electrons in the crystal and also between this pair and the nearest neighboring electron pairs, two of which are in the As$^{3-}$ coordination polyhedron and two are in two neighboring polyhedra located in the second half of the FeAs layer (Fig 7). All of these interactions between electron pairs can also be considered as interactions between corresponding electron crystals they form. Thus, formation of a pair in the electron crystal of unpaired electrons results in a relatively local perturbation in the balance of Coulomb interactions between electron crystals in the sandwich structure of the FeAs layer. However, when the concentration of such perturbations becomes critical (or Coulomb interaction between electron crystals reaches a critical value), a new balance of Coulomb interactions arises and results in a collective change in the state of all unpaired electrons: singlet pairs are formed by unpaired electrons throughout the entire 2D crystal of unpaired electrons, i.e., at all As$^{3-}$, but not only at those As$^{3-}$ which have got additional electrons. Thus, the Coulomb interaction enhancement inside the crystal of unpaired electrons (which form, as shown, triplet pairs with each other) and also with neighboring electron crystals of pairs can result in substitution of triplet pairing in As$^{3-}$ electron shells by singlet pairing. As a consequence, a 2D crystal of singlet electron pairs replaces a 2D crystal of unpaired electrons (or, in another representation, a 2D crystal of triplet pairs). Note that the pairs formed are localized at As$^{3-}$ and become their unshared electron pairs. Since there is a Coulomb interaction between the unshared pairs, the 2D crystal formed by them can be regarded as a 2D Wigner crystal of unshared pairs (see Sect. 4). 

It should also be noted that a doped material can have the amount of unpaired electrons equal to the amount of electrons introduced under doping. (These electrons are the second unpaired electron of each As$^{3-}$ that has got an additional electron under doping). Probably, these electrons will also be localized at anions, but their energy state will differ from the state of unshared pairs. The presence of such electrons can give rise to a weak paramagnetism of the doped material in its normal state, which is observed experimentally \cite{ref10}. The concentration of these unpaired electrons should be maintained at rather a low level not to upset the balance of Coulomb interactions between charged planes in the sandwich-type structure. Thus, there are a lower (critical concentration of perturbations) and upper (critical concentration of paramagnetic electrons) levels of dopant concentration at which superconductivity arises, and this can explain the dome-shaped dependence of Tc on dopant concentration.

\section{Change in the valence electron state under pressure, singlet pairing of unpaired electrons, and formation of electron crystals from the singlet pairs}
\label{se:7}
The materials based on iron pnictides, in contrast to cuprates, proved to be rather sensitive to pressure. It was experimentally shown that application of pressure from 0.35 to 0.86 GPa to CaFe$_2$As$_2$ that contains the same FeAs layers as LaOFeAs can decrease the temperature of transition to the magnetically ordered phase and even induce superconductivity with Tc = 12K without doping \cite{ref19}. Superconductivity in this case arises in the tetragonal phase differing from the high-temperature tetragonal phase, as commonly believed, only by crystal lattice parameters. Thus, superconductivity in FeAs layers can be induced either by introducing additional electrons at doping or a change in crystal lattice parameters. Since doping, as shown, is accompanied by changing of the state of unpaired electrons, the question arises as to whether crystal lattice compression under pressure can also lead to a change in the unpaired electron state.

Before answering this question, let us see why the materials based on iron pnictides are so sensitive to the hydrostatic pressure. A high sensitivity can be caused by the crystal structure of the FeAs layer consisting of FeAs$_4$ tetrahedra and the nonequivalence of chemical bonds in them. Really, shrinkage of a weaker 1e bond must be much more substantial than that of a 2e bond the shrinkage of which can be neglected in the pressure interval typically used. Moreover, nonequivalent bonds are arranged so that the layer thickness is equal to the sum of their projections on the Z axis (Fig.2). Under these circumstances the FeAs layer shrinkage at hydrostatic pressure is bound to be highly anisotropic and to occur mainly along the Z axis via As$^{3-}$ displacements towards Fe$^{2+}$ planes in both parts of the FeAs layer (Fig.8). Just this shrinkage anisotropy was observed experimentally \cite{ref19}. It can be seen from Fig.8 that shrinkage must be accompanied by a decrease in angle $\beta$ between 1e (or ionic bonds substituting them) and 2e bonds in the Fe$^{2+}$ coordination polyhedron, which was also observed experimentally \cite{ref19}.

The decrease in angle $\beta$ results in an increase in Coulomb repulsion between unpaired electrons and electron pairs of 2e bonds in the Fe$^{2+}$ polyhedra (Fig.8) or, in general, between their electron crystals. This increases the energy of unpaired electrons and, probably, can lead to their transition from the $d^4s$ state of As$^{3-}$ into, for instance, a higher-energy $d^4p$ state of As$^{3-}$ and from the triplet pairing to a singlet one. As a result, a singlet electron pair appears in one of two As$^{3-}$ states occupied earlier by two unpaired electrons, and this pair becomes a new unshared pair of As$^{3-}$. Just like at doping, the unshared pair formation is accompanied by transformation of 2D crystals of unpaired electrons into 2D crystals of unshared pairs (Fig.6b) in each half of the FeAs layer.

\begin{figure*}
  \includegraphics[width=0.75\textwidth]{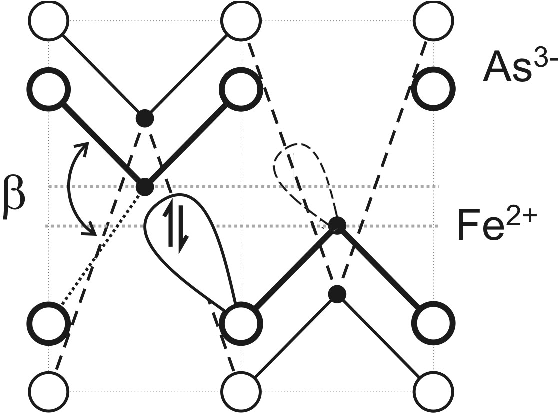}
\caption{Deformation of two neighboring Fe$^{2+}$ tetrahedra under pressure (As$^{3-}$ displacement, a decrease  in angle $\beta$, and formation of an unshared pair in the As$^{3-}$ state and Fe$^{2+}$ unoccupied state) resulting from substitution of 1e bonds by ionic ones. As$^{3-}$ displaced along the Z axis are shown by heavy circles}
\label{fig:8}       
\end{figure*}

\section{Formation of a 2D Wigner crystal of bosons and a unique state of the FeAs layer}
\label{se:8}
As shown above, enhancement of Coulomb interaction between unpaired electrons in the electron crystal they form and/or between the electron crystal of unpaired electrons and crystals of electron pairs (at doping or under pressure) results in singlet pairing of the unpaired electrons and transformation of their 2D electron crystal into a 2D crystal of unshared pairs. This occurs in both parts of the FeAs layer, so there are two 2D crystals of unshared pairs in each FeAs layer. One of them is shown in Fig.9a. The crystals are characterized by a plane-square lattice with the square side equal to the distance (d) between the nearest As$^{3-}$ lying in one and the same plane ($\sim$ 3.9 \AA) and are displaced by $d/\sqrt{2}$ with respect to each to other along the X and Y axes. The unshared pair energy is determined by the As$^{3-}$ electron state energy and is quantized because the As$^{3-}$ electron states are quantized. Due to Coulomb interaction between the unshared pairs, each 2D crystal of unshared pairs can probably be regarded as a 2D Wigner crystal of unshared pairs. It is also important that each of these two 2D Wigner crystals is divided into 1D Wigner crystals since the unshared pairs belong to As$^{3-}$ of separate chains of covalently bonded ions in each half of the FeAs layer (Fig.7).

\begin{figure*}
  \includegraphics[width=0.75\textwidth]{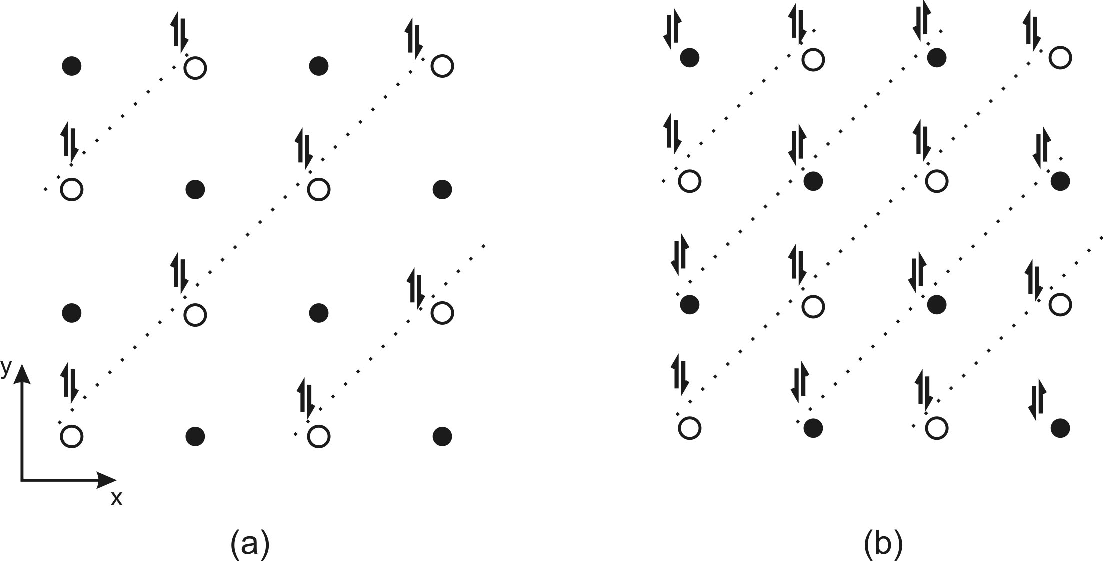}
\caption{Projections (on the XY plane) of 2D Wigner crystal of unshared pairs (prior to the united crystal formation) in one half of the FeAs layer (a) and of the united 2D Wigner crystal of bosons (b) together with projections of two planes of Fe$^{2+}$ ions shown by black and empty circles. The dotted lines show the orientation of the 1D Wigner crystals constituting these 2D Wigner crystals. Different orientations of the double arrows designating bosons in (b) show that they belong to different ion chains} 
\label{fig:9}       
\end{figure*}

As noted above, electron pairing is also accompanied by the appearance of unoccupied hybrid Fe$^{2+}$ states which are partly in the space between the Fe$^{2+}$ planes (Fig.8). These states are oriented approximately in the same direction as the electron states of the Wigner crystal of unshared pairs, i.e., along one of lateral edges of the As$^{3-}$ coordination pyramid. As a result, the unoccupied Fe$^{2+}$ states and the states of unshared pairs are nearly parallel to each other and alternate with each other along the entire ion chain length (Fig.7). 

As temperature decreases and, hence, the crystal lattice of the FeAs layer shrinks, the Coulomb interaction between the unshared pairs and pairs of 2e bonds in the As$^{3-}$ coordination pyramid increases and forces the unshared pairs (or, in general, the 2D Wigner crystal of the unshared pairs) to move from the As$^{3-}$ planes towards the Fe$^{2+}$ planes. In all probability, when the 2D Wigner crystal consisting of 1D crystals falls into the space between the Fe$^{2+}$ planes which contains unoccupied Fe$^{2+}$ states, lateral or $\pi$ overlapping of ion states (unoccupied Fe$^{2+}$ states and As$^{3-}$ states occupied by unshared pairs, supposedly, $sp^3$ and $d^4p$) along the whole length of each chain of covalently bonded ions can occur. As a result, a delocalized $\pi$ orbital is formed along each chain, which is a quasi-1D electron state extending along the chain. As a consequence, all unshared pairs of each 1D Wigner crystal turn out to be in exactly the same state. In other words, the unshared pairs of the 1D (and, hence, 2D) Wigner crystal are converted into bosons. Thus, the ion state overlapping gives rise to boson formation. Due to hybridization of ion states at their overlapping, the boson energy becomes higher than the unshared pair energy. This energy increase is likely to compensate for the bonding energy of the unshared pair with an individual As$^{3-}$. As a result, the bosons are shared by all ions (As$^{3-}$ and Fe$^{2+}$) in the chain, i.e., bonding to an individual As$^{3-}$ is replaced by bonding to ions of the entire chain. Such bosons can be delocalized along the chain without any resistance if two conditions are satisfied. The conditions are as follows (i) bosons can be delocalized only as parts of the 2D Wigner crystal of bosons and (ii) Coulomb interactions between the 2D Wigner crystal of bosons and all charged planes in the FeAs layer must compensate each other. 

It is quite probable that the ion state overlapping in the space between Fe$^{2+}$ planes is accompanied by coalescence of two (from both parts of the FeAs layer) 2D Wigner crystals of bosons and formation of a united 2D Wigner crystal (Fig.6c).The united crystal is characterized by a square-plane boson arrangement with a distance between the nearest bosons of d/$\sqrt{2}$, where d is the distance between unshared pairs in the initial crystals (Fig. 9b). The decrease in the distances is accompanied by increasing Coulomb interaction both between the bosons inside the united 2D Wigner crystal and between the 1D Wigner crystals constituting the united 2D crystal and belonging alternately to the ion chains of one and the other halves of the FeAs layer.

In addition, the united 2D Wigner crystal that lies between the Fe$^{2+}$ planes proves to be in a symmetric (with respect to the charged planes in the sandwich structure) surrounding of charged Fe$^{2+}$ planes, crystals of electron pairs of 2e bonds, and As$^{5+}$planes (charge displacements in the XY plane are ignored). In this position (Fig. 6c) the Coulomb interactions between the united 2D Wigner crystal and all the charged planes compensate each other, i.e., there is nothing to inhibit delocalization of the united 2D Wigner crystal of bosons as a whole in the plane of the crystal. 

The united 2D Wigner crystal formation adequately explains, first, the stability of superconductivity which, as shown, is 1D in its nature and, second, the bell-shaped dependence of Tc on the As$^{3-}$ height above the Fe$^{2+}$ planes \cite{ref20}. The height characterizes the FeAs layer shrinkage and, hence, the Coulomb interaction energy in it that determines the mid-position of the united 2D Wigner crystal of bosons in the sandwich structure at which compensation of Coulomb interactions occurs. 

It should also be noted that the united Wigner crystal formation can also occur prior to ion state overlapping (not only simultaneously, as considered above). In this case the bad-metal conductivity preceding superconductivity can be explained by tunneling of the united Wigner crystal of unshared pairs through the gaps between ion states.

Thus, the FeAs layer is characterized by (i) a well-defined sandwich-type charge distribution; (ii) strong Coulomb interactions between all charged planes of the sandwich structure; (iii) a rather high transparency of positively charged ion planes to Coulomb interactions between 2D electron crystals; (iv) quantized electron crystal energy levels; and (v) delocalization of 2D Wigner crystal of bosons (under doping and pressure) without resistance and without any change in its state (which is similar, to a certain extent, to orbital motion of electrons in an atomic shell). All these features point to a unique state of the FeAs layer and make it similar to a macroscopic quantum system.

\section{Conclusion}
\label{se:9}

Thus, the origin of the high-temperature superconductivity in the compounds based on iron pnictides is adequately described by a chemical model involving additional chemical bonding between ions realized by 1D Wigner crystals of electron pairs and delocalized $\pi$ orbitals (similar to cuprates). 

Owing to studies of these new materials, a deeper insight into specific features of high-Tc superconductivity which have not been revealed in cuprates has been achieved.  Specifically, it has been convincingly shown that superconductivity results from a strongly correlated electron state of all valence electrons, which manifests itself as formation of electron crystals. The origin of the strongly correlated electron state is strong Coulomb interactions between ions (polarization of anions by cations) and between electrons (including electron pairs).

The strongly correlated electron state makes the FeAs layer similar to a macroscopic quantum system which is a quasi-2D negatively charged structure as if "suspended" between positively charged layers having another chemical composition. It is limited by a thickness of a few {\AA} along the Z axis (about 8 {\AA} in iron pnictides) and has an unlimited size in the XY plane. The macroscopic quantum system is characterized by:

\begin{enumerate}
\item a charge ordering of the sandwich type, i.e., alternation of charged planes of 2D electron crystals (which are at different quantized energy levels) and positively charged ion planes;
\item a strong Coulomb interaction between all the charged planes;
\item a rather high transparency of positively charged ion planes to Coulomb interaction between 2D electron crystals lying on both sides of the ion plane (or planes). The transparency results from appreciably differing distances between the charged planes and between charges in each plane and also from displacements of charges in neighboring planes. 
\end{enumerate}

Superconductivity in such a system is accomplished by a 2D Wigner crystal consisting of 1D Wigner crystals of electron pairs condensed into one and the same quasi-1D state, i.e., the pairs converted into bosons. The state is formed by lateral overlapping of ion states along the entire length of the covalently bonded ion chain and extends along the chain. Delocalization of the 2D boson Wigner crystal occurs in the case of compensation of all Coulomb interactions between the crystal and all charged planes which can result from doping, pressure application, and/or temperature decrease. 

Since properties of such a macroscopic quantum system are fully dominated by electron crystals and Cou\-lomb interactions between them, i.e., by strong collective interactions, the system is able to resist thermal vibrations of crystal lattice (and even affect these vibrations and make them strongly anisotropic). This explains why superconductivity in such systems can occur at rather high temperatures. 

To summarize,  the consideration of new superconducting materials based on iron pnictides contributed to a further development of the qualitative model of the high-temperature superconductivity.  In spite of the fact that this model is only qualitative and originates from logical speculations inferred from experimental data,  it shows that strong Coulomb interaction between ions and between electrons leads to a highly correlated electron state and, under certain conditions (doping, pressure application and/or temperature decrease) transform the FeAs layer into the state similar to a macroscopic quantum system. Superconductivity in such a system is accomplished by a two-dimensional Wigner crystal consisting of one-dimensional Wigner crystals formed by bosons, i.e., singlet electron pairs that are in the same quasi-one-dimensional state extending along the ion chain. The correctness of the model  is confirmed by the fact that it is able to adequately explain salient properties of three different classes of superconducting materials (cuprates, doped picene, iron pnictides) and their changes under the action of external factors. Therefore, it can be supposed that the model can prove useful for development of the theory of superconductivity taking into account the highly correlated state of valence electrons and strong Coulomb interactions between these electrons.
\section{Acknowledgements}
The author would like to acknowledge very useful discussions on characteristic properties of superconductivity in materials with strongly correlated electron state with Academician Yu.V. Kopaev and Prof. K.D.Tsendin

\bibliographystyle{model1-num-names}




\end{document}